\documentclass[10pt, conference, letterpaper]{IEEEtran}

\usepackage{graphicx} 
\usepackage[utf8]{inputenc}
\usepackage[T1]{fontenc}
\usepackage{microtype}
\usepackage{xargs}
\usepackage{lipsum}
\usepackage{xparse}
\usepackage{xifthen, xstring}
\usepackage{xspace}
\usepackage{marginnote}
\usepackage{etoolbox}
\usepackage{soul}
\usepackage{cite}
\usepackage{amsmath,amsfonts}
\usepackage{algorithmic}
\usepackage{graphicx}
\usepackage{textcomp}
\usepackage{hhline}
\usepackage{multirow}
\usepackage{booktabs}
\usepackage{subcaption}
\usepackage{tabularray}
\usepackage{pifont}
\usepackage[dvipsnames]{xcolor}
\usepackage{float}
\usepackage{colortbl}
\usepackage{threeparttable}
\usepackage{tablefootnote}
\usepackage{adjustbox}

\usepackage{amssymb}
\usepackage{pifont} 
\usepackage{newtxtext}
\usepackage{xcolor}
\usepackage[colorlinks=black, urlcolor=black, linkcolor=black]{hyperref}
\definecolor{Rosewater}{RGB}{220, 138, 120}
\definecolor{Flamingo}{RGB}{221, 120, 120}
\definecolor{Pink}{RGB}{234, 118, 203}
\definecolor{Mauve}{RGB}{136, 57, 239}
\definecolor{Red}{RGB}{210, 15, 57}
\definecolor{Maroon}{RGB}{230, 69, 83}
\definecolor{Peach}{RGB}{254, 100, 11}
\definecolor{Yellow}{RGB}{223, 142, 29}
\definecolor{Green}{RGB}{64, 160, 43}
\definecolor{Teal}{RGB}{23, 146, 153}
\definecolor{Sky}{RGB}{4, 165, 229}
\definecolor{Sapphire}{RGB}{32, 159, 181}
\definecolor{Blue}{RGB}{30, 102, 245}
\definecolor{Lavender}{RGB}{114, 135, 253}

\usepackage{tabularx}
\usepackage{multicol}

\usepackage{stfloats}

\newcommand{\xdmaname}{\mbox{\textit{XDMA}}\xspace}
\newcommand{\xdmathreename}{\mbox{\textit{XDMA}3}\xspace}
\newcommand{\xdmafivename}{\mbox{\textit{XDMA}5}\xspace}
\newcommand{\xdmaninename}{\mbox{\textit{XDMA}9}\xspace}
\newcommand{\xdmasname}{\mbox{\textit{XDMA}s}\xspace}

\title{\xdmaname: A Distributed, Extensible DMA Architecture for Layout‑Flexible Data Movements in Heterogeneous Multi-Accelerator SoCs}

\date{November 2025}

\begin{document}

\author{
    \IEEEauthorblockN{Fanchen Kong$^*$, Yunhao Deng$^*$, Xiaoling Yi, Ryan Antonio, Marian Verhelst}
    \IEEEauthorblockA{
        MICAS-ESAT, KU Leuven, Belgium \\
        \{fanchen.kong, yunhao.deng, xiaoling.yi, ryan.antonio, marian.verhelst\}@esat.kuleuven.be
    }
}

\maketitle

\def\thefootnote{}\footnotetext{This project has been partly funded by the European Research Council (ERC) under grant agreement No. 101088865, the European Union’s Horizon 2020 program (CONVOLVE) under grant agreement No. 101070374, the Flanders AI Research Program, and KU Leuven.
 }\def\thefootnote{\arabic{footnote}}

\def\thefootnote{*}\footnotetext{Both authors contributed equally to this research.}\def\thefootnote{\arabic{footnote}}

\begin{abstract}
As modern AI workloads increasingly rely on heterogeneous accelerators, ensuring high-bandwidth and layout-flexible data movements between accelerator memories has become a pressing challenge. Direct Memory Access (DMA) engines promise high bandwidth utilization for data movements but are typically optimal only for contiguous memory access, thus requiring additional software loops for data layout transformations. This, in turn, leads to excessive control overhead and underutilized on-chip interconnects. To overcome this inefficiency, we present \xdmaname, a distributed and extensible DMA architecture that enables layout-flexible data movements with high link utilization. We introduce three key innovations: (1) a data streaming engine as \xdmaname Frontend, replacing software address generators with hardware ones; (2) a distributed DMA architecture that maximizes link utilization and separates configuration from data transfer; (3) flexible plugins for \xdmaname enabling on-the-fly data manipulation during data transfers. \xdmaname demonstrates up to 151.2$\times$/8.2$\times$ higher link utilization than software-based implementations in synthetic workloads and achieves 2.3$\times$ average speedup over accelerators with SoTA DMA in real-world applications. Our design incurs $<$2\% area overhead over SoTA DMA solutions while consuming 17\% of system power. \xdmaname proves that co-optimizing memory access, layout transformation, and interconnect protocols is key to unlocking heterogeneous multi-accelerator SoC performance.
\end{abstract}


\section{Introduction}
\label{sec:introduction}

The growing demand for compute performance and advances in silicon technology have driven the integration of multiple heterogeneous accelerators into single Systems-on-Chip (SoCs)\cite{esp_2024_isscc,nectar_ra_soc_2024_hc} to achieve higher performance and energy efficiency in compute-intensive tasks. These accelerators include Generalized Matrix-Matrix Multiplication (GeMM) accelerators \cite{davinci_2019_HCS}, In-Memory Computing (IMC)\cite{axelera_dimc_2024_isscc}, 
accelerators for sparse data\cite{shi2024bitwave}, and security coprocessors\cite{security_acc_cosic_2023_jssc}. To achieve high energy efficiency and avoid stalls, these accelerators often employ dedicated memory subsystems. In practice, however, while data access between memory subsystems and accelerators is heavily optimized, the data exchange across different accelerators is overlooked, limiting the overall performance of heterogeneous SoCs. Copying data across heterogeneous accelerators presents two interrelated challenges: (1) Modern workloads are increasingly memory-bounded due to a lack of data reuse (2) The in-memory data layout must align with the accelerators' diverse access patterns, 
such as a tiled layout for GeMM, a row-major layout for SIMD, etc. Suboptimal layouts can increase inference latency up to 100$\times$ compared to optimal accelerator-tailored formats\cite{feather_2024_isca} because explicit data layout transformations are costly in terms of energy and latency. 

\begin{figure}[tp]
    \centering
    \includegraphics[width=0.95\linewidth]{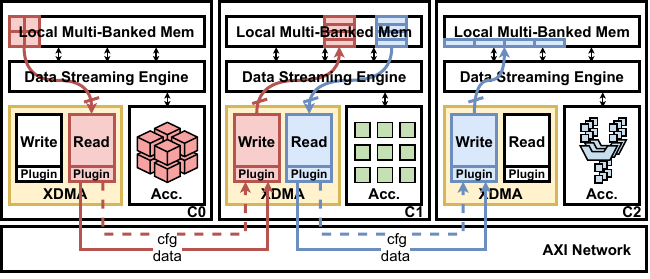}
    \caption{\xdmaname moves data in a multi-accelerator SoC}
    \label{fig:overall}
\end{figure}

Direct Memory Access (DMA) engines are key components for achieving high-bandwidth data movements between memories. However, traditional DMAs can only copy contiguous data sequences. Thus, layout transformation is only achievable through software control loops, which incurs significant control overhead. A possible mitigation involves offloading layout transformations to standalone accelerators, allowing DMAs to retain burst transfers. However, this approach introduces additional latency and energy costs for intermediate data, undermining the benefits of accelerator specialization. To overcome this inefficiency, we propose \textbf{\xdmaname}\footnote{\href{https://github.com/KULeuven-MICAS/snax_cluster/tree/main/hw/chisel/src/main/scala/snax/xdma}{\textbf{\underline{\xdmaname Frontend}}} is open-sourced as one component in \href{https://github.com/KULeuven-MICAS/snax_cluster}{\textbf{{\underline{SNAX}}}}, while \href{https://github.com/KULeuven-MICAS/xdma_axi_adapter}{\textbf{\underline{\xdmaname Backend}}} is open-sourced separately. } (the operating mechanism is shown in Fig. \ref{fig:overall}), an extensible DMA architecture that unifies high-utilization memory transfers and efficient data layout transformation. 

The main \textbf{contributions} of this work are: 
\begin{itemize}
    \item We design a distributed DMA architecture with decoupled read/write ports, communicating through a two-phase circuit-switched protocol, bypassing AXI limitations to sustain high link utilization. 
    \item We replace software-managed loops by a hardware solution, enabling N-dimensional affine address generation with minimal bandwidth penalties.
    \item We design standardized and flexible \xdmaname plugins to support on-the-fly compute and layout transformation. 
    \item We validate \xdmaname using synthetic and real workloads, achieving 8.2$\times$ to 151.2$\times$ improvements vs. software loop baselines, and on average 2.7$\times$ higher link utilization vs. accelerator+DMA design. \xdmaname incurs less than 2\% area overhead compared to SoTA DMA, 17\% of system energy.
\end{itemize}

Table \ref{tab:sota_comparison} compares the \xdmaname architecture with SoTA DMAs from academia and industry.

\begin{table*}[tp]
  \centering
  \footnotesize
  \begin{tabularx}{0.95\textwidth}{lllllll}
    \toprule
    & \textbf{Architecture}
    & \textbf{Technology}
    & \textbf{Address Gen}
    & \textbf{Data Access}
    & \textbf{Comp-while-transfer}
    & \textbf{Open-Sourced} \\
    \midrule
    HyperDMA\cite{hyperdma_2024_icicm}
      & \textcolor{ForestGreen}{Distributed}
      & \textcolor{Maroon}{RTL}
      & \textcolor{ForestGreen}{ND}
      & \textcolor{YellowOrange}{Direct (Coarse-grained)}
      & \textcolor{Maroon}{None}
      & \textcolor{Maroon}{No} \\
    
    ESP DMA\cite{esp_2024_isscc}
      & \textcolor{YellowOrange}{Monolithic}
      & \textcolor{YellowOrange}{FPGA}
      & \textcolor{Maroon}{1D}
      & \textcolor{Maroon}{Through interconnect}
      & \textcolor{Maroon}{None}
      & \textcolor{ForestGreen}{Yes} \\
    
    Gemmini DMA\cite{gemmini-dac}
      & \textcolor{YellowOrange}{Monolithic}
      & \textcolor{ForestGreen}{FPGA, Silicon}
      & \textcolor{YellowOrange}{2D}
      & \textcolor{Maroon}{Through interconnect}
      & \textcolor{YellowOrange}{Transpose, Scaling}
      & \textcolor{ForestGreen}{Yes} \\
    
    IDMA\cite{idma_paper}
      & \textcolor{YellowOrange}{Monolithic}
      & \textcolor{ForestGreen}{FPGA, Silicon}
      & \textcolor{ForestGreen}{Optional ND}
      & \textcolor{Maroon}{Through interconnect}
      & \textcolor{YellowOrange}{In-stream Acc. port}
      & \textcolor{ForestGreen}{Yes} \\
    
    AMD DMA v7.1\cite{amd_dma}
      & \textcolor{YellowOrange}{Monolithic}
      & \textcolor{YellowOrange}{FPGA}
      & \textcolor{YellowOrange}{Optional 2D}
      & \textcolor{Maroon}{Through interconnect}
      & \textcolor{Maroon}{None}
      & \textcolor{Maroon}{No} \\
    
    TI EDMA3\cite{ti_dma}
      & \textcolor{YellowOrange}{Monolithic}
      & \textcolor{YellowOrange}{Silicon}
      & \textcolor{YellowOrange}{3D}
      & \textcolor{Maroon}{Through interconnect}
      & \textcolor{Maroon}{None}
      & \textcolor{Maroon}{No} \\
    
    \midrule
    \textbf{\xdmaname}
      & \textbf{\textcolor{ForestGreen}{Distributed}}
      & \textbf{\textcolor{ForestGreen}{FPGA, Silicon}}
      & \textbf{\textcolor{ForestGreen}{ND}}
      & \textbf{\textcolor{ForestGreen}{Direct (Fine-grained)}}
      & \textbf{\textcolor{ForestGreen}{Flexible Plugins}}
      & \textbf{\textcolor{ForestGreen}{Yes}} \\
    
    \bottomrule
  \end{tabularx}
  \caption{Comparison between \xdmaname and SoTA DMA architectures}
  \label{tab:sota_comparison}
\end{table*}
\section{\xdmaname Architecture}
\label{sec:architecture}
\begin{figure*}[tp]
\vspace{-1.5em}
    \centering
    \includegraphics[width=0.85\linewidth]{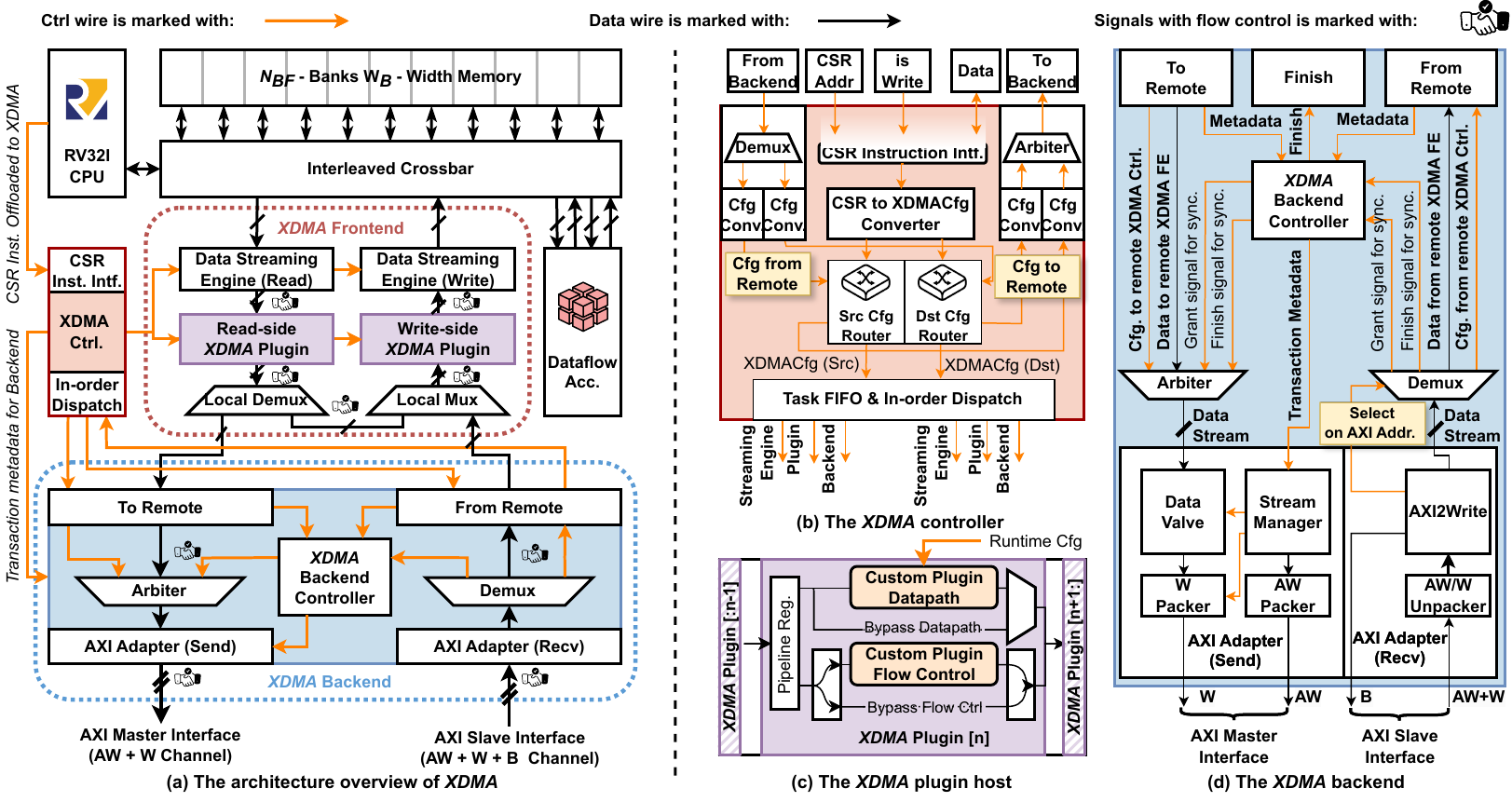}
    \caption{The architecture of \xdmaname(a) and three selected sub-modules (b, c, d)}
    \label{fig:archi_main_figure}
\end{figure*}

\begin{table}[tp]
    \centering
    \begin{threeparttable}
        \footnotesize
        \begin{tabularx}{\linewidth}{llll}
            \toprule
            \textbf{Parameters} & \textbf{Symbol} & \textbf{Parameters} & \textbf{Symbol} \\
            \midrule
            Mem. Base Addr. & $Addr_{Mem}$ & Mem. Width & $W_{B}$ \\
            Mem. Size & $Size_{Mem}$ & AXI Width & $W_{AXI}$ \\
            Src./Dst. Buf. Depth & $D_{Buf,src/dst}$ & Src./Dst. Dim. & $Dim_{src/dst}$\\
            Src./Dst. \#Channel & $N_{C,src/dst}$ & Src./Dst. Ext. List & $Ext_{src/dst}$\\
            \bottomrule
        \end{tabularx}
        \caption{Design-time parameters of \xdmaname}
        \label{tab:param}
    \end{threeparttable}
\end{table}

\xdmaname proposes a novel decentralized DMA architecture. Fig. \ref{fig:archi_main_figure}(a) shows the hardware hierarchy of \xdmaname. The \xdmaname Controller (\S\ref{subsec:archi_controller}) receives instructions and forwards configurations to local or remote \xdmaname unit. The \xdmaname Datapath (\S\ref{subsec:archi_datapath}) includes three building blocks: (1)
The Frontend interfaces with the memory, offering flexible memory accesses; (2) The Plugin provides a standardized interface for integrating customized modules that manipulate data during transfers; (3) The Backend encapsulates data into AXI frames and manages the tunnel between two half-\xdmasname units. \xdmaname's parameterized design enables easy integration into various SoCs. Table \ref{tab:param} details \xdmaname’s design-time parameters.
\subsection{\xdmaname Orchestration}
\label{subsec:archi_protocol}
\begin{figure}[tp]
    \centering
    \includegraphics[width=0.9\linewidth]{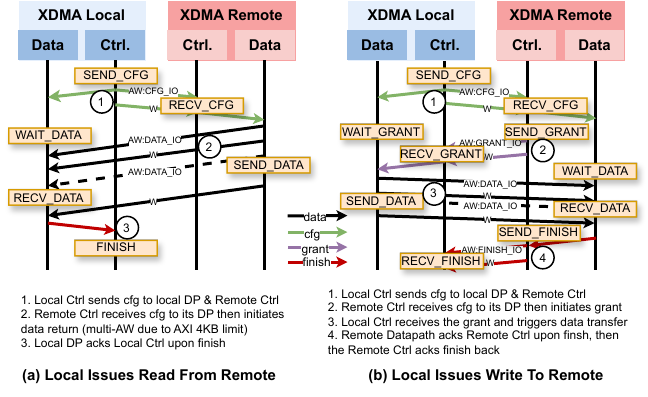}
    \caption{The \xdmaname orchestration for read and write requests}
    \label{fig:archi_protocol}
\end{figure}

The coordination between two half \xdmasname consists of a two-phase flow following the circuit switching principle, as depicted in Fig. \ref{fig:archi_protocol}: first, a \textbf{CFG transfer phase}, where transaction CFG are forwarded to the remote counterpart, and then a \textbf{Data transfer phase} in which the link is fully occupied by data. 
\xdmaname employs a distributed architecture where each unit features both master and slave ports, distinguished from the conventional DMAs that only use master ports. This enables \xdmaname to encapsulate all transfers into AXI write request, simplifying the design while maintaining full-duplex transfer capability. Although CFG and Data transactions share the AW/W channel, deadlock will not occur as all transactions are one-to-one, and no dependency between different \xdmaname transactions. Furthermore, \xdmaname can autonomously arbitrate the task based on the FIFO principle if contention happens. 

\subsection{\xdmaname Controller}
\label{subsec:archi_controller}
\xdmaname Controller (Fig. \ref{fig:archi_main_figure}(b)) converts the offloaded CSR instruction into \texttt{XDMACfg} structures to describe one \xdmaname task.  Then, two configuration routers route the \texttt{XDMACfg} to \xdmaname that attached to the correct memory region. Two routers forward the CFG to remote side through the AXI interconnect if this task needs the collaboration of two \xdmaname. Finally, Src. and Dst. configurations arrive at the Task FIFO and In-order Dispatch unit, which monitors the status of the Frontend, and dispatches a new task when the previous one is finished. 

\subsection{\xdmaname Datapath}
\label{subsec:archi_datapath}
The \xdmaname Frontend (shown in Fig. \ref{fig:archi_main_figure}(b)) accesses local memories in an N-D affine pattern and prefetches the data for the \xdmaname Backend. We utilize data streaming engines designed for dataflow accelerators \cite{yi2025datamaestro}, consisting of a $Dim$-dimensional address generator and a $D_{buf}$-depth data buffer. The address generator effectively offloads the address computation tasks from the processor, and the data buffer mitigates potential bank conflicts during transformation of diverse data layouts. 

The \xdmaname Backend (Fig. \ref{fig:archi_main_figure}(d)) acts as the compatibility layer between the Frontend and the AXI4 interconnect. It establishes \textit{virtual tunnels} between two \xdmasname units on top of the AXI protocol by mapping signals (\textit{cfg}, \textit{data}, \textit{grant}, \textit{finish}) to independent MMIO addresses and initiate write requests to counterpart's MMIOs. Each Backend is both AXI Master and Slave, so every two pairs can collaborate independently. 

\xdmaname enables on-the-fly data manipulation during local and remote data transfers through custom Plugins that can be inserted within the \xdmaname Frontend. Two Plugin Hosts, one post-reader and one pre-writer, share a uniform architecture, as depicted in Fig. \ref{fig:archi_main_figure}(c). One or more plugins can be cascaded and each plugin can have its own control bit vectors.

\section{\xdmaname Evaluation}
\label{sec:evaluation}

We evaluate the performance of \xdmaname in a multi-accelerator SoC.  We first analyze the ability of \xdmaname to efficiently transform data layouts (\S\ref{subsec:evaluation_reshape}). To validate \xdmaname's adaptability to real workloads, we subsequently prototype this \xdmaname-attached SoC on the AMD Versal\textsuperscript{\texttrademark} VPK180 FPGA and demonstrate its efficacy in KV-cache prefill/load workloads for the DeepSeek-V3\cite{liu2024deepseek} LLM (\S\ref{subsec:evaluation_fpga}). Finally, we synthesize the design into an ASIC implementation to obtain the area and power results (\S\ref{subsec:evaluation_asic}). 
We vary $D_{buf,src/dst}$, a key design-time parameter affecting performance-area trade-offs, from 3 to 9, to showcase how this parameter impacts the \xdmaname design. 

\subsection{System Environment Setup}
\label{subsec:evaluation_setup}

We group a 4MB, 32-bank, 64-bit-per-bank memory, two RV32I cores\cite{snitch_2020}, an accelerator, and an \xdmaname into an accelerator cluster. Since this paper does not focus on evaluating accelerator's performance, we attach a basic 8$\times$8$\times$8 GeMM mainly for the area estimation. We setup a dual-cluster SoC derived from Occamy\cite{paulin2024occamy} to evaluate \xdmasname, the width of AXI interconnect is configured as 512 bits. \textit{DataMaestro}\cite{yi2025datamaestro} is chosen as the data streaming engine of \xdmaname. Our baselines are the iDMA\cite{idma_paper} and Gemmini's built-in DMA\cite{gemmini-dac}, which can represent SoTA general-purpose DMA and workload-optimal DMA respectively.  

All RTL simulations are conducted using Verilator. Silicon synthesis is performed using Synopsys Design Compiler\textsuperscript{\textregistered} with GF 22nm FDX\textsuperscript{\texttrademark} technology at 1GHz 0.8V. Power consumption is analyzed using the synthesized netlist and gate-level switching activity via Synopsys PrimeTime\textsuperscript{\textregistered}.

\subsection{Matrix Layout Transformation}
\label{subsec:evaluation_reshape}
\begin{figure}[tp]
    \centering
    \includegraphics[width=0.85\linewidth]{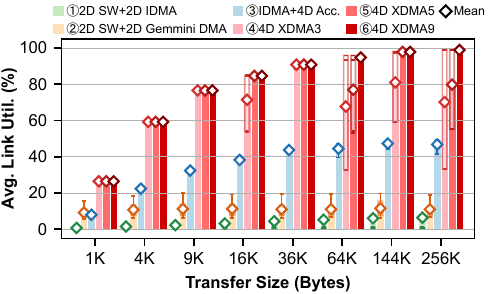}
    \caption{The average link utilization for the 4D matrix reshape: \ding{172} 2D software control loop + 2D iDMA copy, \ding{173} 2D software control loop + 2D Gemmini DMA copy, \ding{174} iDMA copy + dedicated 4D layout transformation accelerator, and \ding{175}\ding{176}\ding{177} 4D \xdmaname with $D_{buf,src/dst}=3,5,9$ (\xdmathreename/5/9)}
    \label{fig:eval_reshape}
\end{figure}

We compare the average link utilization of various 4D data layout transformation and data copy workloads across different HW/SW setups. We select four data layouts: MN, MNM8N8, MNM8N16 and MNM8N32, which is the optimal data layout for 2D/3D GeMM array. The matrix size is chosen from 32$\times$32 to 512$\times$512. Six different HW/SW setups are evaluated as shown in Fig. \ref{fig:eval_reshape}, resulting in total 768 test points. 

The effective BW of each test is calculated by the total data volume transferred divided by the measured execution time. The link utilization is then calculated by dividing the effective BW by the theoretical BW. For software-managed DMA setups (\ding{172},\ding{173}), the address calculations occur before data movement. 

The results show that \xdmaninename (\ding{177}) achieves the highest and most stable link utilization. All hardware-accelerated solutions (\ding{174}-\ding{177}) outperform software-loop approaches (\ding{172}, \ding{173}) by a wide margin, confirming our claim in \S\ref{sec:introduction} that software control can create performance bottlenecks. Among software-managed approaches, the Gemmini\cite{gemmini-dac} DMA outperforms the iDMA\cite{idma_paper} due to the higher I/O performance of the Rocket core, reducing control overhead. The solution \ding{174} improves on the former two, but incurs additional memory overheads due to intermediate results. In general, \xdmaninename (\ding{177}) exceeds SoTA solutions (\ding{172}\ding{173}\ding{174}) by 151.2$\times$/8.2$\times$/2.4$\times$ on average, respectively. 

When comparing \xdmaname with different $D_{buf}$ (\ding{175}-\ding{177}), \xdmaninename (\ding{177}) outperforms \xdmathreename and \xdmafivename by 1.7$\times$ and 1.1$\times$ on average. We observe higher variations for the setups with the smaller $D_{buf}$ because the smaller buffer cannot consistently prevent stalls caused by bank conflicts. All remaining tests are conducted on \xdmaninename for maximum performance. 

\subsection{Real Workload Evaluation on FPGA}
\label{subsec:evaluation_fpga}

\begin{figure}[tp]
    \centering
    \begin{minipage}{0.32\linewidth}
        \centering
        \includegraphics[width=\linewidth]{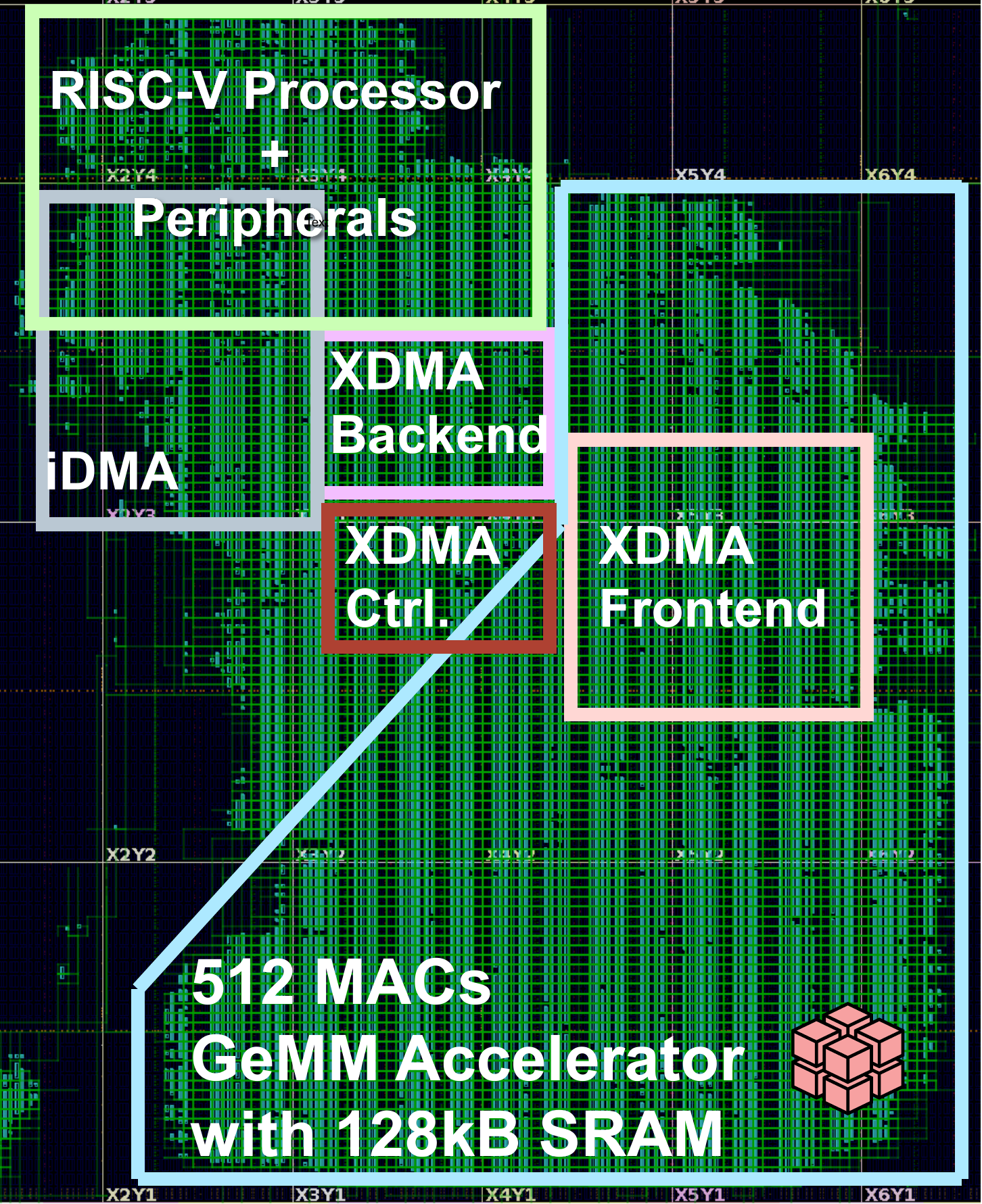}
    \end{minipage}
    \begin{minipage}{0.45\linewidth}
        \centering
        \begin{threeparttable}
            \footnotesize
            \begin{tabular}{c|c}
                \toprule
                \textbf{Platform}            & Versal\textsuperscript{\texttrademark} VPK180  \\
                \textbf{Frequency}           & 100MHz         \\
                \hline
                \textbf{LUTs Total}          & 268k           \\
                \textbf{Regs Total}          & 67k            \\
                \textbf{LUTs GeMM}           & 150k  (56.0\%) \\
                \textbf{Regs GeMM}           & 18k   (26.9\%) \\
                \textbf{LUTs iDMA}           & 8k     (2.9\%) \\
                \textbf{Regs iDMA}           & 8.8k  (13.1\%) \\
                \textbf{LUTs \xdmaninename}  & 20.5k  (7.6\%) \\
                \textbf{Regs \xdmaninename}  & 5.9k   (8.8\%) \\
                \bottomrule
            \end{tabular}
        \end{threeparttable}
    \end{minipage}
    \caption{The FPGA result of the accelerator cluster with \xdmaname}
    \label{fig:fpga_impl}
\end{figure}

We implement the evaluating clusters on a VPK180 FPGA. Fig. \ref{fig:fpga_impl} details the annotated FPGA floorplan, operating frequency, and resource utilization for an 8$\times$8$\times$8 GeMM accelerator cluster featuring with \xdmaname, showing that \xdmaname introduces approximately 8\% area overhead.

Next, we benchmark cross-cluster data copy performance offered by \xdmaname using Deepseek-v3’s KV-cache matrix shape of
8192$\times$512 with Batch=1, representing personal-use scenarios. Evaluated workloads include: (1) Prefill stage: a GeMM accelerator in cluster 1 (Optimal layout: MNM8N8) computes the KV cache, followed by an RMSNorm on a SIMD accelerator in cluster 2 (Optimal layout: MN). Finally, the RMSNormed data is stored to another cluster in the MNM8N8 layout; (2) Load stage: The KV-cache data is after a first GeMM in cluster 1, transferred and simultaneously transposed to cluster 2 for transformer operations. Table \ref{tab:kvcache} shows that the workloads executing on the \xdmaname experience a 2.3$\times$ latency improvement compared with the baseline setup (iDMA+Accelerator).

\begin{table}[tp]
\begin{tabular}{@{}c@{\hspace{4pt}}|c@{\hspace{4pt}}c@{\hspace{4pt}}c|@{\hspace{4pt}}r@{}}
\toprule
\textbf{Experiment} & \textbf{Shape}   & \textbf{I/O Layout} & \textbf{Operation} & \textbf{\#CC/Acc. ratio} \\
\hline
Prefill 1& 2048$\times$512 & MNM8N8/MN          & Reshape     & 38542 / 2.34$\times$     \\
Prefill 2& 2048$\times$512 & MN/MNM8N8          & Reshape     & 42934 / 2.60$\times$     \\
\hline
Load 1   & 2048$\times$512 & MNM8N8             & Transpose   & 37509 / 2.28$\times$   \\
Load 2   & 4096$\times$512 & MNM8N8             & Transpose   & 74884 / 2.28$\times$   \\
Load 3   & 8192$\times$512 & MNM8N8             & Transpose   & 149639 / 2.28$\times$   \\
\bottomrule
\end{tabular}%
\caption{KV-Cache Prefill/Load evaluation of \xdmaname}
\label{tab:kvcache}
\end{table}

\subsection{Area and Power Evaluation}
\label{subsec:evaluation_asic}
\begin{figure}[tp]
    \centering
    \includegraphics[width=\linewidth]{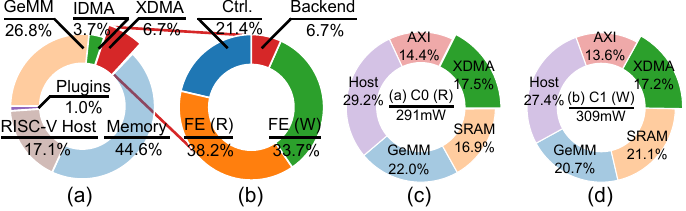}
    \caption{Area and power decomposition of \xdmaname evaluation cluster: (1)(2) The area breakdown; (3)(4) The power breakdown when data being copied from cluster 0 to cluster 1.}
    \label{fig:eval_area_power}
\end{figure}

Finally, we synthesize the cluster with 8$\times$8$\times$8 GeMM and an \xdmaname (with a reduced memory size of 128kB) into an ASIC implementation. \xdmaname occupies 6.7\% of the accelerator cluster's area and consumes 17\% of the total cluster power (Fig. \ref{fig:eval_area_power}) when executing 1D memory copy task. 

\section{Conclusion}
\label{sec:conclusion}
In this paper, we present \xdmaname, a distributed and extensible DMA architecture designed for efficient memory layout transformations across heterogeneous accelerators. \xdmaname offers 8.2$\times$-151.2$\times$ improvements over the SW-managed solutions in non-contiguous data copy tasks. We also present implementation results on both FPGA and silicon technology. \xdmaname incurs less than 2\% area overhead compared to iDMA and consumes 17\% of the total power of the accelerator cluster. 
\bibliographystyle{ieee_trans_single_author.bst}
\bibliography{refs}

\end{document}